\documentclass[11pt]{article}
\usepackage[sc]{mathpazo}
\usepackage{fullpage}
\usepackage[authoryear,sectionbib,sort]{natbib}
\usepackage{graphicx}
\usepackage{adjustbox}
\usepackage[boxruled,linesnumbered]{algorithm2e}
\usepackage{multirow}
\usepackage{amsmath}
\usepackage[utf8]{inputenc}
\usepackage{lineno}
\usepackage{titlesec}
\usepackage{graphicx}
\usepackage{subcaption}
\usepackage{booktabs,multirow}
\usepackage{here}
\usepackage{url}

\newcommand{\mat}[1]{\mbox{\boldmath{$#1$}}} 

\usepackage{listings}
\usepackage{xcolor}

\lstdefinestyle{mystyle}{
	backgroundcolor=\color{backcolour},   
	commentstyle=\color{codegreen},
	keywordstyle=\color{magenta},
	numberstyle=\tiny\color{codegray},
	stringstyle=\color{codepurple},
	basicstyle=\ttfamily\footnotesize,
	breakatwhitespace=false,         
	breaklines=true,                 
	captionpos=b,                    
	keepspaces=true,                 
	numbers=left,                    
	numbersep=5pt,                  
	showspaces=false,                
	showstringspaces=false,
	showtabs=false,                  
	tabsize=2
}

\definecolor{codegreen}{rgb}{0,0.6,0}
\definecolor{codegray}{rgb}{0.5,0.5,0.5}
\definecolor{codepurple}{rgb}{0.58,0,0.82}
\definecolor{backcolour}{rgb}{0.95,0.95,0.92}

\titleformat{\section}[block]{\Large\bfseries\filcenter}{\thesection}{1em}{}
\titleformat{\subsection}[block]{\Large\itshape\filcenter}{\thesubsection}{1em}{}
\titleformat{\subsubsection}[block]{\large\itshape}{\thesubsubsection}{1em}{}
\titleformat{\paragraph}[runin]{\itshape}{\theparagraph}{1em}{}[. ]

\title{Modelling longitudinal polytomous animal data using Bayesian hierarchical models}

\author{Maria Letícia Salvador$^{1\ast}$ \and 
Gabriel R. Palma$^{2, 3}$ \and
Mariana Coelly Modesto Santos Tavares$^{4}$\and
Iran José Oliveira Silva$^{4}$ \and
Idemauro Antonio Rodrigues de Lara$^{1}$
}

\date{}

\begin{document}
\maketitle
\noindent{} 1. University of São Paulo,  Luiz de Queiroz College Agriculture,  Exact Sciences Department, Piracicaba, São Paulo, Brazil;

\noindent{} 2. Hamilton Institute, Maynooth University, Maynooth, Ireland;

\noindent{} 3. Department of Mathematics and Statistics, Maynooth University, Maynooth, Ireland;

\noindent{} 4. University of São Paulo, Luiz de Queiroz College Agriculture, Biosystems Engineering, Piracicaba, São Paulo, Brazil;

\noindent{} $\ast$ Corresponding author; e-mail: marialesalvador@gmail.com

\bigskip


\bigskip


\bigskip

\textit{Manuscript type}: Research paper. 

\bigskip

\noindent{\footnotesize Prepared using the suggested \LaTeX{} template for \textit{Am.\ Nat.}}

\newpage{}

\section*{Abstract}
The analysis of longitudinal categorical data can be complex and unfeasible due to the number of parameters involved, characterised by overparameterisation leading to model non-convergence, in addition to problems related to sample size and the presence or absence of overdispersion. In this context, we introduce Bayesian hierarchical models as an alternative methodology to classical statistical techniques for analysing nominal polytomous data in longitudinal studies. The theoretical foundation is based on the use of non-informative priors and advanced computational techniques, such as Markov Chain Monte Carlo (MCMC) methods, which enable a robust and flexible data analysis framework. As a motivating example, the procedure is illustrated through an applied study in agrarian science, focusing on animal welfare, which assessed seven types of behaviours exhibited by pigs over twelve weeks. The results demonstrated the efficacy of Bayesian hierarchical models for the analysis of longitudinal nominal polytomous data. Since the computational procedures were implemented in the R software and the codes are available, this work will serve as support for those who need such analyses, especially in agricultural designs, where longitudinal categorical data are frequently encountered.

\newpage{}

\section{Introduction} 
The importance of animal welfare is widely recognised in fields such as agricultural science, ethics, and animal production, reflecting the understanding that animal welfare influences not only the health and quality of life of animals but also the quality of animal-derived products and, consequently, human health \citep{mkwanazi2019effects}. In this context, studies focused on animal welfare are fundamental, driving the development and application of advanced statistical methodologies for rigorous and reliable analyses.

Research on pig welfare has highlighted the significance of environmental enrichment through the use of toys and other objects to alleviate stress, stimulate growth, promote health, and improve the animals’ zootechnical indicators \citep{li2021effects, dos2021preference}. Furthermore, environmental enrichment has positively impacted behavioural development and the learning capacity of pigs, enhancing activities such as tail wagging and playing, which are indicative of welfare \citep{ralph2018enrichment}. Therefore, implementing environmental enrichment in pig rearing systems represents an effective strategy to promote animal welfare, benefiting both the animals and producers \citep{giuliotti2019effect}.


A crucial aspect of analysing data from animal welfare studies is the nature of the response variables involved. These variables are often nominal polytomous, meaning they are categorised without a natural order among categories. Frequently, these variables are observed over time in longitudinal studies, where the associated distribution is multinomial. However, analysing such data requires specific statistical techniques, as methods designed for continuous or ordinal variables may not be appropriate. A common approach for analysing nominal polytomous response variables in longitudinal contexts is the use of generalised mixed logits models \citep{hartzel2001multinomial, agresti2005bayesian, chan2023multilevel}, in which one category is treated as the reference.

A recurring challenge in using these models is the potential for parameter estimation problems using the maximum likelihood method, especially in cases involving more than three response categories and inadequate sample sizes. As an alternative, the Bayesian approach employs non-informative priors, allowing for a more flexible and adaptive analysis of the data. This approach eliminates the need for restrictive prior assumptions, enabling the data to have a more direct influence on parameter inferences. Additionally, Bayesian estimation requires computationally intensive programming, such as Monte Carlo methods via Markov Chains (MCMC), particularly the Metropolis-Hastings algorithm and the Gibbs sampler~\citep{gelman1995bayesian}. These are the most commonly used methodologies for obtaining summaries of the posterior joint distribution, which forms the basis of Bayesian inference. In this context, Bayesian hierarchical models for nominal polytomous data \citep{natarajan2000, pettitt2006, gong2004} emerge as a promising alternative. By incorporating a prior structure and utilising advanced computational techniques, these models enable efficient parameter estimation even in the presence of overparameterisation or complex data.

Within this framework, this study proposes Bayesian hierarchical models as an alternative to classical statistical methods, employing non-informative priors and the MCMC method. Although Bayesian methods are known in the statistical literature, they are still emerging in the analysis of longitudinal categorical data. Thus, it is expected that this work can contribute to the dissemination and use of such procedures, especially in agricultural designs.

\section{Materials and Methods}

\subsection{Material}

As a motivating study, we considered an experiment conducted by \cite{tavares2023enriquecimento} in a commercial pig slaughterhouse in Brazil between April and August 2019. The primary objective was to assess the impact of various environmental enrichment conditions on pig behaviour.

\begin{figure}[H]
    \centering
    \includegraphics[width=1\textwidth]{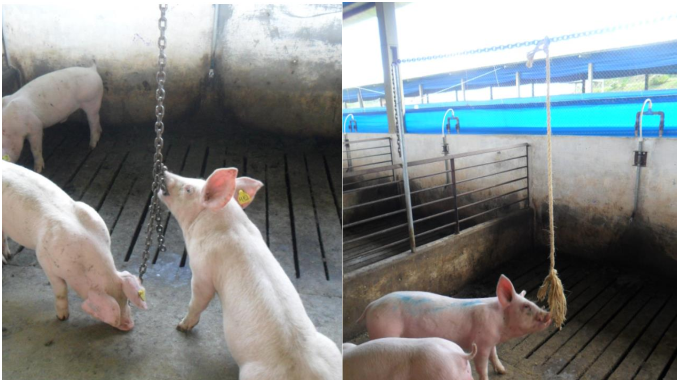}
    \caption{Branched chain and branched sisal rope environmental enrichments were inserted into the pens during the experiment conducted by Tavares (2023) between April and August 2019.}
    \label{fig:CA}
\end{figure}

The experiment consisted of a randomized block design, that corresponds the three weight pigs groups: light ($20.3$ to $24.8$ kg), medium ($24.8$ to $27.7$ kg), and heavy ($27.7$ to $33.8$ kg), with a $2 \times 3$ factorial scheme ($2$ classifications according to the animal sex and $3$  environmental enrichment types: branched chains, branched sisal ropes, and no enrichment) involving $432$ pigs of both sexes, aged between $63$ and $70$ days, distributed in $72$ pens, each containing $6$ animals. 

For each combination of enrichment conditions and sex, 12 animals were selected, designated as T1 - branched chain for males, T2 - branched sisal rope for males, T3 - an environment without enrichment for males, T4 - branched chain for females, T5 - branched sisal rope for females, and T6 - an environment without enrichment for females.

The response variables included behaviors such as (1) aggressiveness, (2) feeding, (3) calm, (4) animal interaction, (5) environmental interaction, (6) object interaction, and (7) locomotion. Animal behaviour observations were conducted by a single observer using the ``Focal Analysis'' method over 12 weeks.

\subsection{Methods}\label{metodo}
\noindent

Before establishing the statistical model, it is always important to conduct an exploratory analysis of the data. For our responses, considering the seven classifications of animal behaviours as nominal polytomous variables, many techniques are multivariate. Here, we have applied the correspondence analysis  \citep{greenacre2010correspondence} by reducing the size of the data and making the appropriate associations with the study factors. Additionally, as is classic with longitudinal data, we have used mean profile plots \cite{singer2018analise} to explore seasonal variation in behaviours across weeks.

We proposed a Bayesian hierarchical model for nominal polytomous response variables. To establish notation, the random variable $\mbox{Y}_{ijt}$ represents the type of behaviour adopted by the $i$-th pig, $i=1, \cdots, 72$, in the $j$-th behaviour classification, $j=1, \cdots, 7$ (aggressive, feeding, calm, animal interaction, environmental interaction, object interaction, and locomotion), evaluated in the $t$-th week, $t=1, \cdots, 12$, following a multinomial distribution, that is, $\mat{Y}_{ijt} \sim \mbox{Multinomial}(y_{ijt}, \pi_{ijt})$. To account for potential dependencies among repeated observations on the same animal, a random effect for pigs (experimental units) is included in the model. By setting ``Aggressive'' behaviour ($J=1$) as the reference category, the Bayesian hierarchical model for polytomous variables is defined as:

\begin{equation} \label{modelcomp}
\begin{split}
     \eta_{ijt} &= \ln \left[\frac{\pi_{ijt}}{\pi_{i1t}}\right] \\
     &= \alpha_{jt} + \beta_{jt} (\mbox{block})_{jt} + \gamma_{jt}(\mbox{sex})_{jt} \\
     &+ \varphi_{jt} (\mbox{enrichment})_{jt} + \omega_{jt} (\mbox{sex} \times \mbox{enrichment})_{jt} + u_i,
\end{split}
\end{equation}

\vspace{0.3cm}

\noindent where $\alpha_{jt}$ is the intercept for the $j$-th classification of pig behaviour collected at the week $t$, $\beta_{jt}(\mbox{weight})_{jt}$ is the effect associated with the pig weight classification, $\gamma_{jt}(\mbox{sex})_{jt}$ the effect of the sex, $\varphi_{jc}(\mbox{enrichment})_{jc}$ the effect associated with the environmental enrichment condition, $\omega_{jt}(\mbox{sex x enrichment})_{jt}$ the interaction effect between enrichment condition and pig sex, and $u_i$ is the random effect associated with the $i$-th observed pig, for which $u_i \sim \mat{N}_n (0, \Sigma_j)$, where $0$ is the mean vector, and $\Sigma_j$ is the variance-covariance matrix.

{ The following model structures were considered:}

\begin{enumerate}
    \item \textbf{Model 1:} considers the block effect with the addition of sex, environmental enrichment condition, and random effects, given by:
    
    \begin{equation} \label{model1}
     \eta_{ijt} = \alpha_{jt} + \beta_{jt} (\mbox{block})_{jt} + \gamma_{jt}(\mbox{sex})_{jt}
     + \varphi_{jt} (\mbox{enrichment})_{jt} + u_i
     \end{equation}

     \item \textbf{Model 2:} considers the interaction effect between pig sex and environmental enrichment conditions with random effects, which is given by:
     \begin{equation} \label{model2}
     \eta_{ijt} = \alpha_{jt} + \beta_{jt} (\mbox{block})_{jt} + \gamma_{jt}(\mbox{sex})_{jt}
     * \varphi_{jt} (\mbox{enrichment})_{jt} + \omega_{jt} (\mbox{sex} \times \mbox{enrichment})_{jt} + u_i
     \end{equation}
\end{enumerate}

Model selection was performed using the deviance information criterion (DIC). Thus, this information criterion was used to check the effect of interaction by selecting the model which produced the lowest DIC. 

The estimation of  models \ref{model1} and \ref{model2} were conducted under the Bayesian approach. To capture the uncertainties of the individual effects $\mat{u}_i$, which represent the specific variations associated with each pig, a normal prior distribution with mean $\mat{0}$ and variance $0.01$ was used. These effects reflect the variability of pigs and the uncertainty associated with their influence on the model. Furthermore, for the parameters $\mat{\alpha}_{j}, \mat{\beta}_{jk}, \mat{\gamma}_{js}, \mat{\varphi}_{jc}$, which can assume any real value, normal prior distributions with mean $\mat{0}$ and variance $0.01$ were employed. These priors address the uncertainty related to the influence of these parameters on the model. Specifically:

\[u_i \sim N(0, 0.01),\]
\[\alpha_j \sim N(0, 0.01),\]
\[\beta_k \sim N(0, 0.01),\]
\[\gamma_s \sim N(0, 0.01) \hspace{0.2cm} \text{and}\]
\[\varphi_c \sim N(0, 0.01).\]

The posterior distribution was obtained using the Markov Chain Monte Carlo (MCMC) method. Four MCMC chains were used, each with $5000$ iterations, followed by $1000$ burn-in iterations at every $15$ steps to assess chain convergence. Implementation was performed using \texttt{JAGS} \citep{hornik2003jags} through the \texttt{rjags} package \citep{rjags}, with \texttt{R} software \citep{R}. 

After the burn-in phase, the analysis proceeded to calculate $95\%$ credibility intervals for the model parameters~\citep{gelman1995bayesian}. These intervals were calculated using the $2.5^{\text{th}}$ and $97.5^{\text{th}}$ percentiles of the posterior distributions of MCMC samples, providing a clear view of the uncertainty associated with the parameter estimates.

\section{Results and Discussion}

\begin{figure}[H]
\centering
\includegraphics[width=0.9\textwidth]{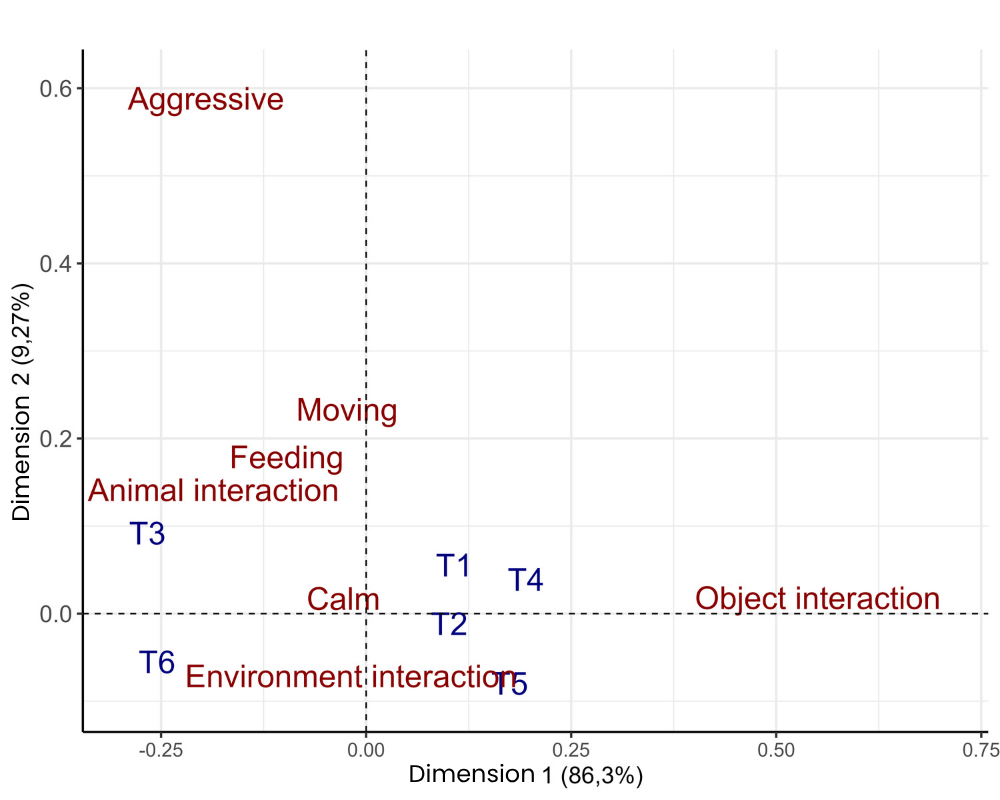}
 \caption{Representative graph of the types of behaviours attributed by pigs to different treatment conditions, according to the experiment conducted by Tavares(2023) between April and August 2019. The treatments are as follows: T1 - branched chain for males, T2 - branched sisal rope for males, T3 - unenriched environment for males, T4 - branched chain for females, T5 - branched sisal rope for females, and T6 - unenriched environment for females.}
 \label{CA}
\end{figure}

Initially, the association between the behaviours of pigs (aggressive, feeding, calm, animal interaction, environmental interaction, object interaction, and locomotion) and the treatments to which they were subjected during the experiment (T1 - branched chain for males, T2 - branched sisal rope for males, T3 - unenriched environment for males, T4 - branched chain for females, T5 - branched sisal rope for females, and T6 - unenriched environment for females) was investigated using the chi-square test ($\chi^2$). The test was significant  (\textit{p-value} $<$ 0.05), indicating evidence of an association between pig behaviours and the treatments, which serves as the premise for correspondence analysis, presented in Figure \ref{CA}, that illustrates this association.

According to the correspondence analysis  (Figure \ref{CA}), there is a $96.1\%$ representation in two dimensions. It was observed that specific treatments, such as T3 (unenriched environment for male pigs), showed a possible trend of association with behaviours such as ``Animal Interaction'', ``Feeding'', and ``Locomotion'', suggesting that male pigs without environmental enrichment have a certain preference for interacting with other animals or moving around. On the other hand, the treatment T6 (unenriched environment for female pigs) appears to be associated with calm behaviour, indicating that, unlike male pigs, females tend to remain calmer when not subjected to environmental enrichment. Moreover, treatments T1 (branched chain for male pigs), T2 (branched sisal rope for male pigs), and T4 (branched chain for female pigs) are seemingly more associated with the ``Object Interaction'' behaviour, suggesting that females have a greater preference for the branched chain. In contrast, males interact with both the branched chain and the sisal rope when subjected to environmental enrichment. However, these indications are merely exploratory.

\begin{figure}[H]
\centering
\includegraphics[width=0.8\textwidth]{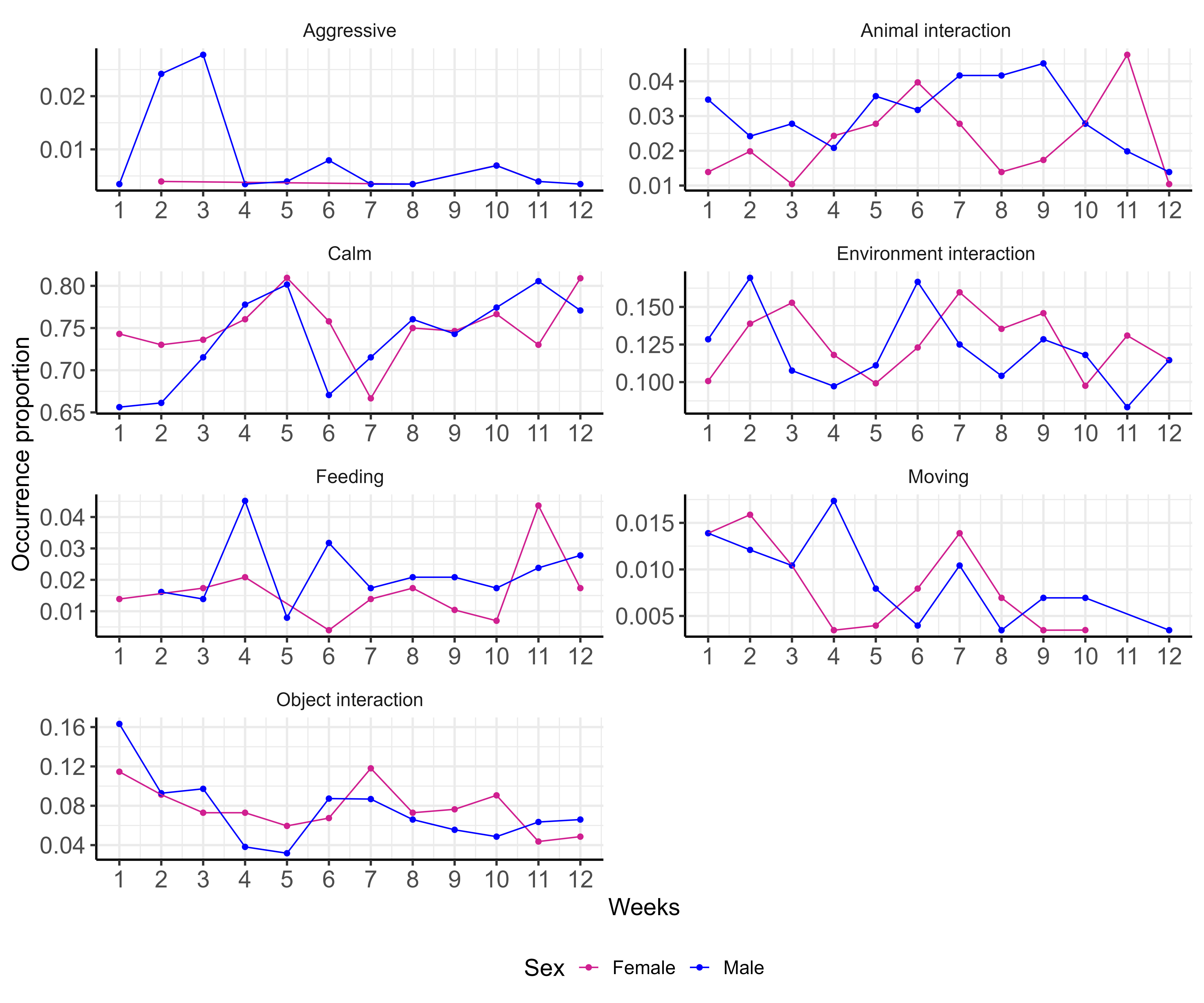}
 \caption{Graph of average profiles for each behavioural classification of pigs over the twelve observed weeks, according to the experiment conducted by Tavares(2023) between April and August 2019.}
 \label{per1}
\end{figure}

The graph of the average profiles for each type of behaviour adopted by the pigs over the twelve weeks is described in Figure \ref{per1}.

Our results show that over the twelve weeks, the ``Calm'' behaviour prevailed among both male and female pigs throughout the experiment. In contrast, behaviours classified as ``Aggressive'', ``Animal Interaction'', and ``Object Interaction'' showed stability in their occurrence, with lower frequencies. Furthermore, the ``Locomotion'' behaviour showed a slight decrease over the weeks.

\begin{figure}[H]
\centering
\includegraphics[width=1.0\textwidth]{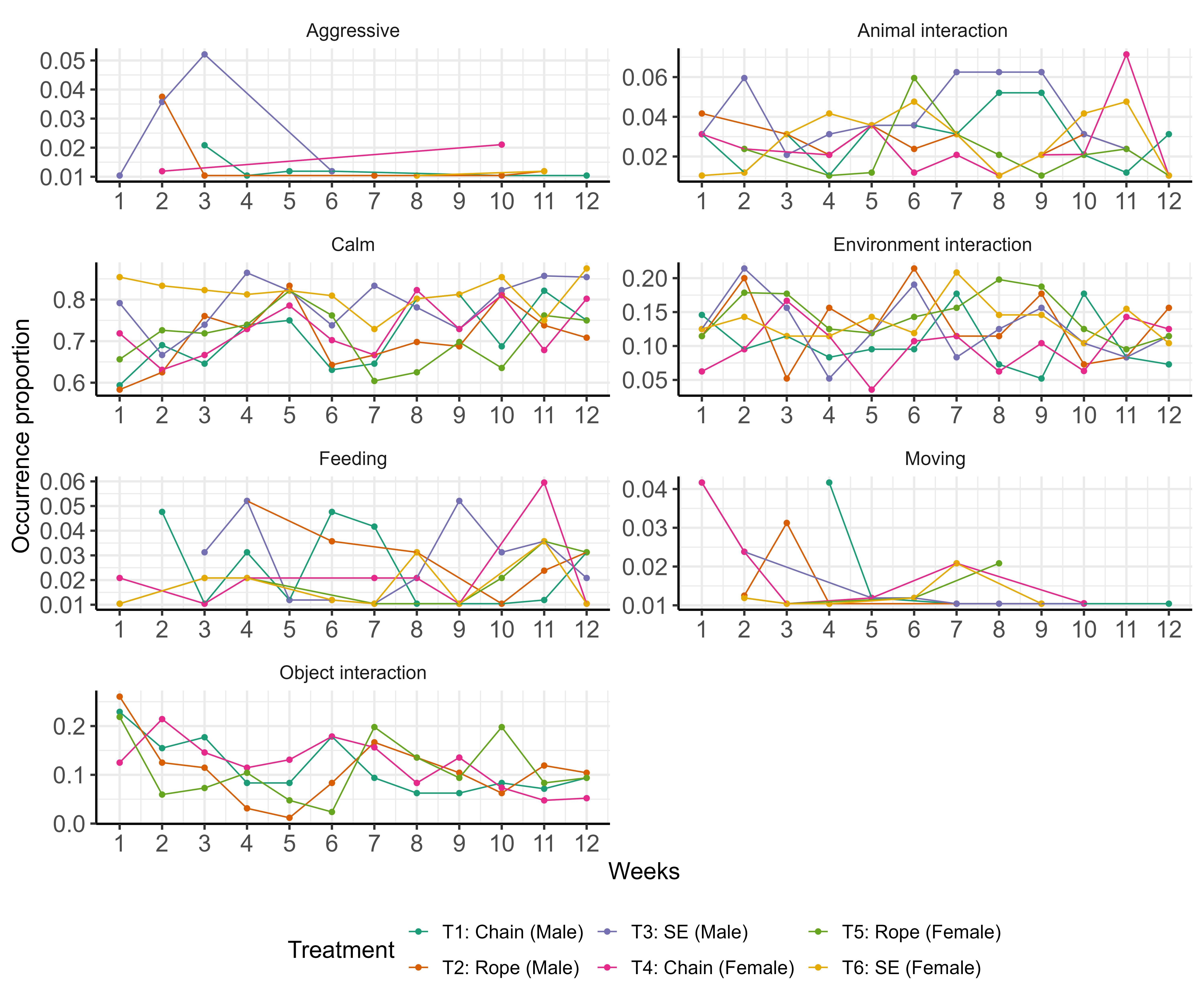}
 \caption{Average profiles for each behavioural classification of pigs under different treatment conditions over the twelve observed weeks, according to the experiment conducted by Tavares(2023) from April to August 2019. The treatments are: T1 - branched chain for males, T2 - branched sisal rope for males, T3 - unenriched environment for males, T4 - branched chain for females, T5 - branched sisal rope for females, and T6 - unenriched environment for females.}
 \label{perf2}
\end{figure}

The proportions of each pig's behaviour for each treatment are presented in Figure \ref{perf2}. It was noted that male pigs appeared to be more aggressive than females at the start of the study. Behaviour categories, including ``Feeding'', ``Animal Interaction'', and ``Environmental Interaction'', showed variations that were not affected by the environmental enrichment condition applied.

Next, Bayesian hierarchical models for nominal polytomous data were fitted, and the selection was made using the deviance information criterion (DIC), as presented in the section \ref{metodo}. The selection result is presented in Table \ref{table:model_selection}.

\captionsetup{justification=justified, singlelinecheck=off,labelfont=bf}
\begin{table}[ht]
\caption{Linear predictor selection using the deviance information criterion (DIC), for the experiment conducted by \cite{tavares2023enriquecimento} between April and August 2019}.
\label{table:model_selection}
\vspace{-0.2cm}
\centering

\begin{tabular}{ccr}
    \hline
Model & Linear predictor                          & DIC     \\ 
\hline
1      & Sex + Enrichment + Random effect & 10158,1 \\
2      & Sex $\times$ Enrichment + Random effect & 10154,5 \\ 
\hline
    \end{tabular}
\end{table}

Through the comparison of the models, Model 2, which incorporates the interaction between sex and enrichment condition, yielded a lower DIC value ($10154.5$), suggesting that this model is more appropriate. The effect of the interaction, in practice, shows that the response to environmental enrichment depends on the level of the sex factor, as illustrated by the Figure \ref{per1}.

\begin{figure}[H]
\centering
\includegraphics[width=1.0\textwidth]{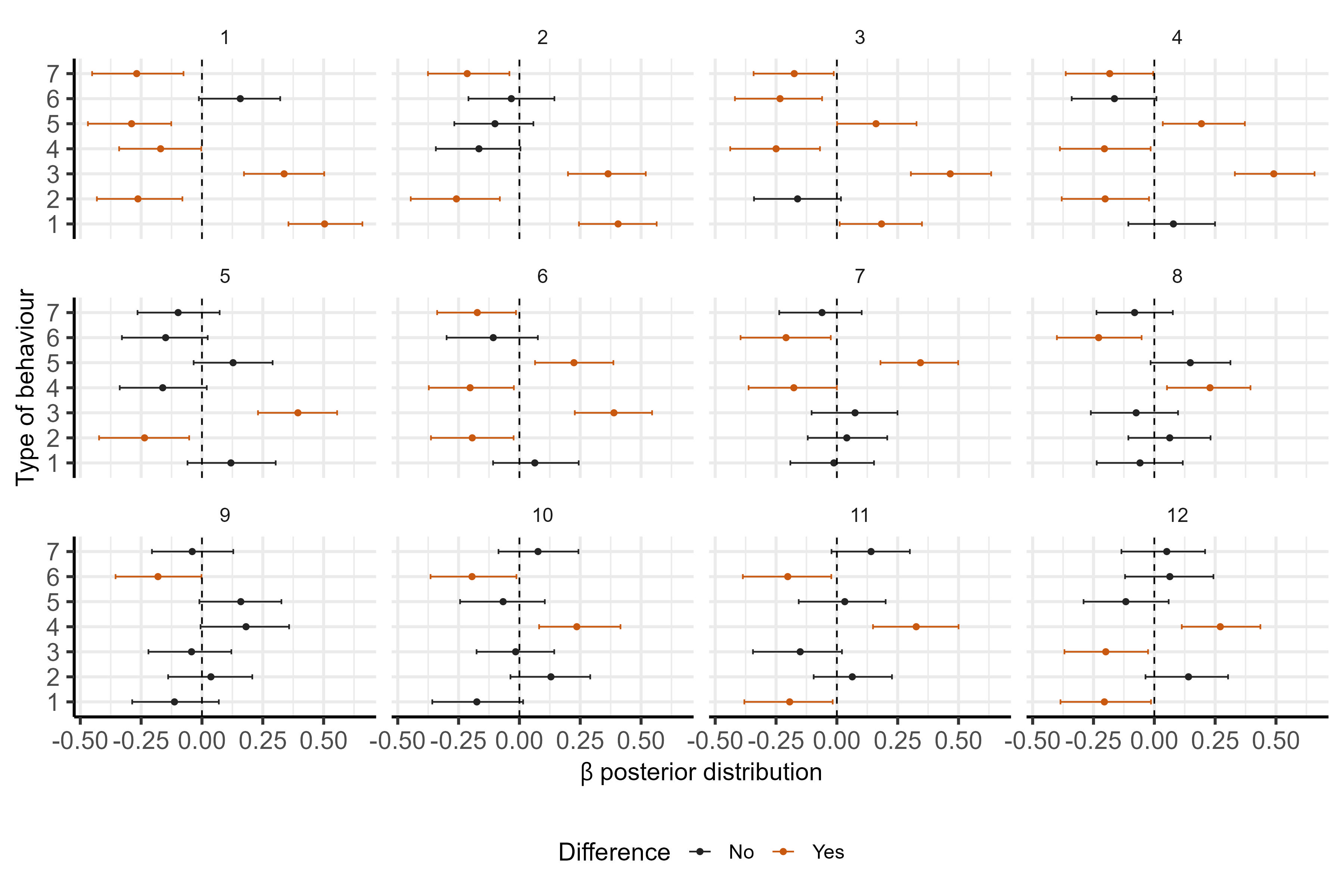}
 \caption{$95\%$ credibility intervals for the estimates of $\beta$, derived from the posterior distribution, indicating the behaviour of female pigs under the environmental enrichment condition with sisal rope. The data were collected over 12 weeks of experimentation, observing the following behaviours: (1) Aggressive, (2) Feeding, (3) Calm, (4) Animal interaction, (5) Environmental interaction, (6) Object interaction, and (7) Locomotion.}
 \label{femea_corda}
\end{figure}

\begin{figure}[H]
\centering
\includegraphics[width=1.0\textwidth]{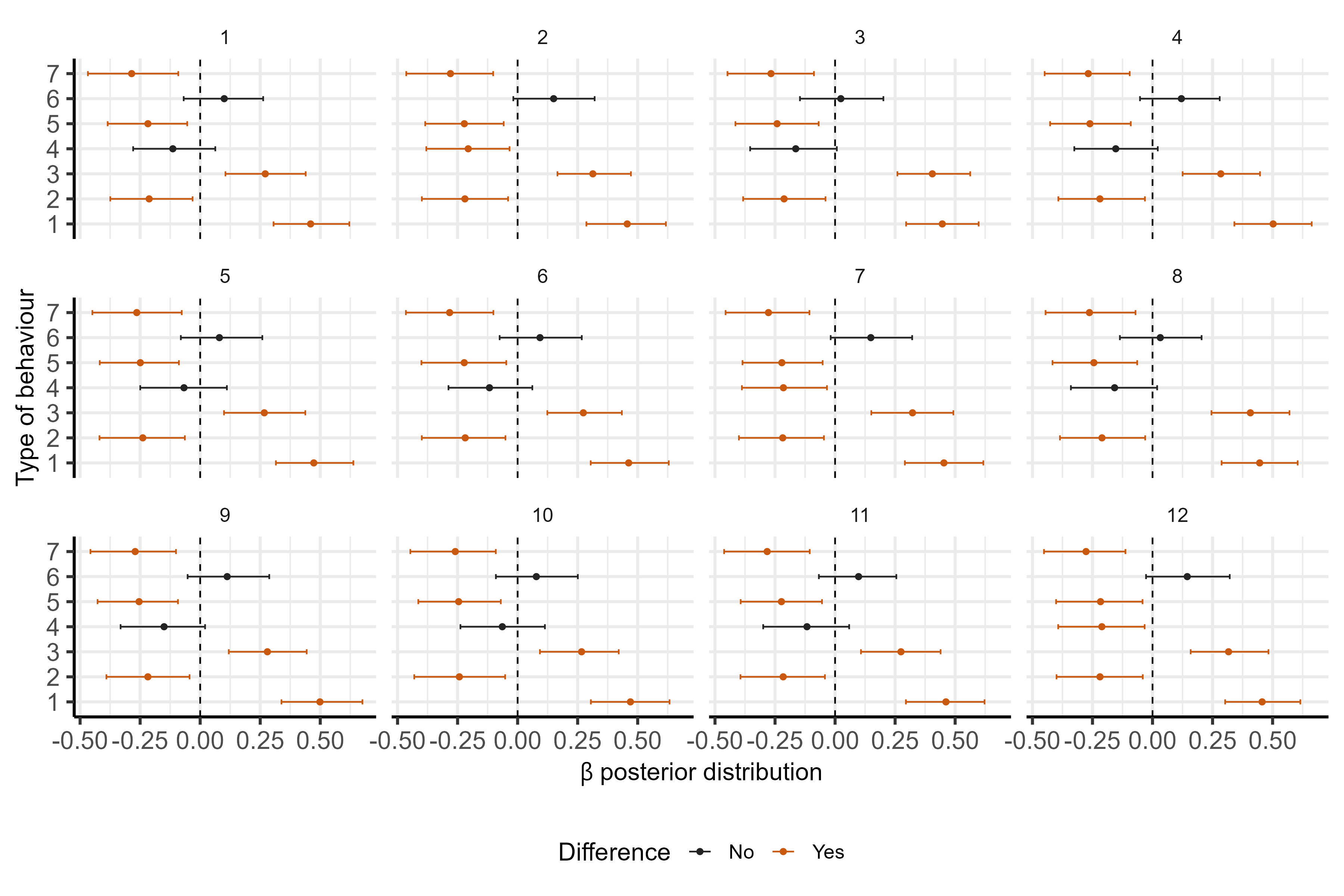}
 \caption{$95\%$ credibility intervals for the estimates of $\beta$, derived from the posterior distribution, indicating the behaviour of female pigs under the environmental enrichment condition with a branched chain. The data were collected over 12 weeks of experimentation, observing the following behaviours: (1) Aggressive, (2) Feeding, (3) Calm, (4) Animal interaction, (5) Environmental interaction, (6) Object interaction, and (7) Locomotion.}
 \label{femea_corrente}
\end{figure}

\begin{figure}[H]
\centering
\includegraphics[width=1.0\textwidth]{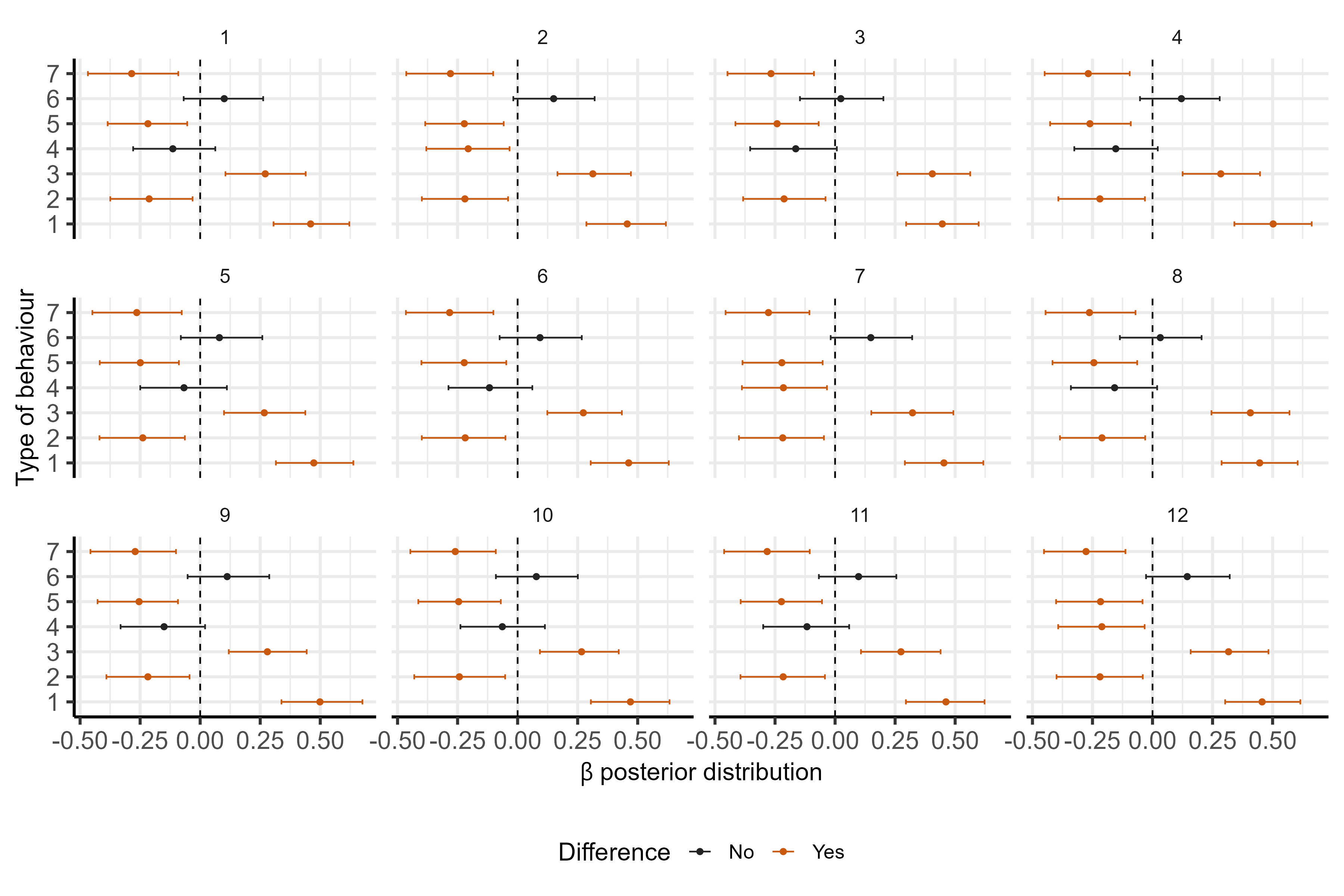}
 \caption{$95\%$ credibility intervals for the estimates of $\beta$, derived from the posterior distribution, indicating the behaviour of female pigs without an environmental enrichment condition. The data were collected over 12 weeks of experimentation, observing the following behaviours: (1) Aggressive, (2) Feeding, (3) Calm, (4) Animal interaction, (5) Environmental interaction, (6) Object interaction, and (7) Locomotion.}
 \label{femea_SE}
\end{figure}

The posterior estimates $\beta$ of the coefficients indicate that environmental enrichment conditions influenced the behaviour of female pigs over the course of $12$ weeks. Overall, the $95\%$ credibility intervals suggest that the ``Aggressive'' behaviour was present in the initial weeks, regardless of the type of enrichment to which the animals were exposed. 

For female pigs exposed to the sisal rope, a transition from ``Aggressive'' to ``Calm'' behaviour was observed from the fourth week onwards. The ``Object Interaction'' behaviour showed significant posterior estimates $\beta$ in week 3 and between weeks 7 and 11, suggesting that the sisal rope attracted interest at specific points during the experiment. However, in the final week, this behaviour was no longer significant, indicating a possible decline in interest in the object (Figure \ref{femea_corda}). Additionally, ``Environmental Interaction'' behaviour was more frequent in week 7, whereas ``Animal Interaction'' increased in weeks 8 and between weeks 10 and 12.

On the other hand, female pigs exposed to the branched chain displayed a consistent pattern of ``Aggressive'' behaviour throughout the experiment. Moreover, ``Object Interaction'' behaviour was not significant in any of the weeks, indicating that this type of enrichment was not appealing to the animals (Figure \ref{femea_corrente}). These results suggest that the sisal rope may have been a more stimulating option for the pigs compared to the branched chain.

In the absence of enrichment, the posterior estimates $\beta$ indicate that ``Aggressive'' behaviour predominated throughout the entire experiment (Figure \ref{femea_SE}). Furthermore, aside from ``Object Interaction'', the only behaviour that did not reach significance was ``Animal Interaction''. In contrast, the behaviours ``Feeding'', ``Environmental Interaction'', and ``Locomotion'' occurred less frequently, suggesting that the lack of stimuli limited behavioural variation and may have been associated with a higher level of stress.

\begin{figure}[H]
\centering
\includegraphics[width=1.0\textwidth]{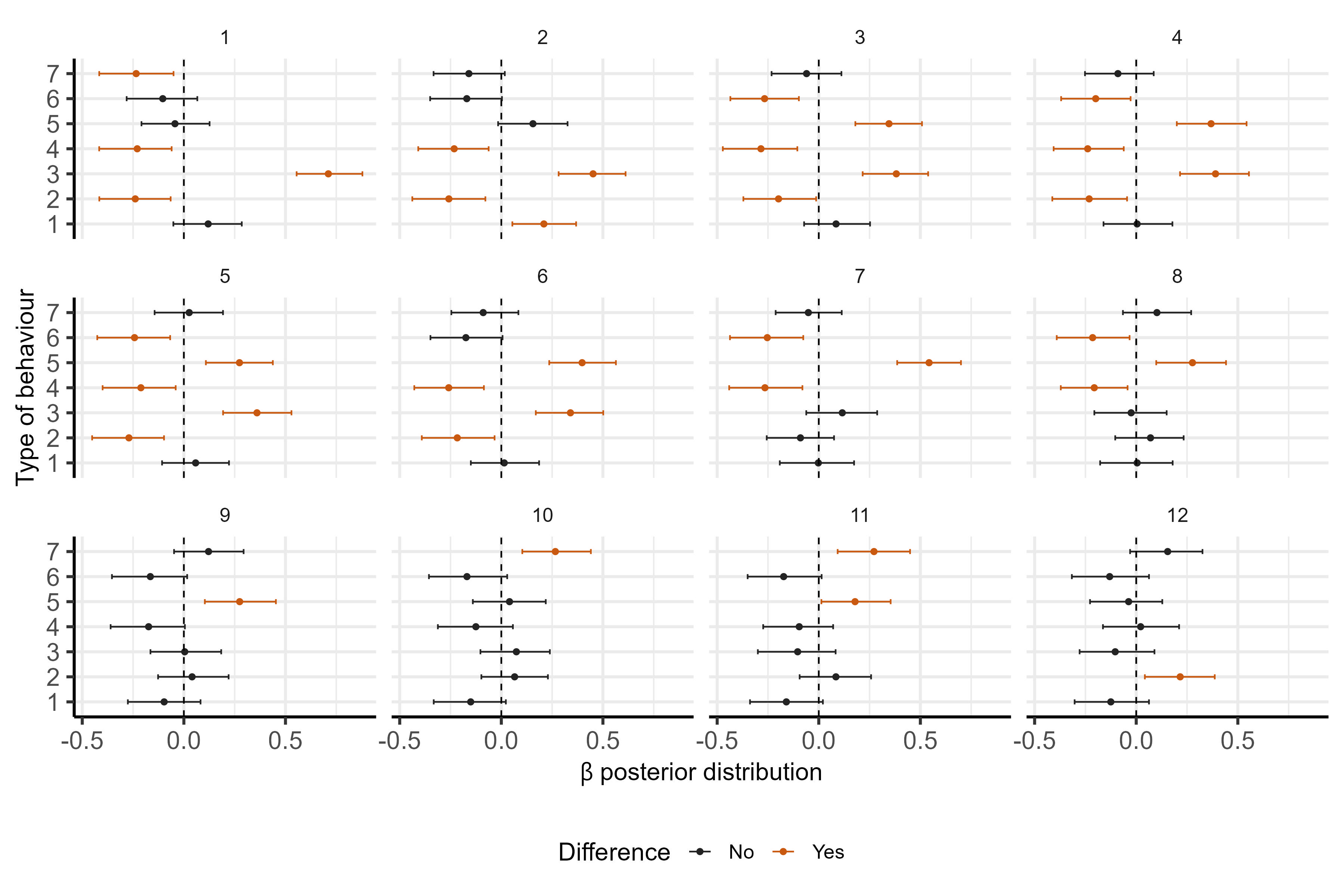}
 \caption{$95\%$ credibility intervals for the estimates of $\beta$, derived from the posterior distribution, indicating the behaviour of male pigs under the environmental enrichment condition with sisal rope. The data were collected over 12 weeks of experimentation, observing the following behaviours: (1) Aggressive, (2) Feeding, (3) Calm, (4) Animal interaction, (5) Environmental interaction, (6) Object interaction, and (7) Locomotion.}
 \label{macho_corda}
\end{figure}

\begin{figure}[H]
\centering
\includegraphics[width=1.0\textwidth]{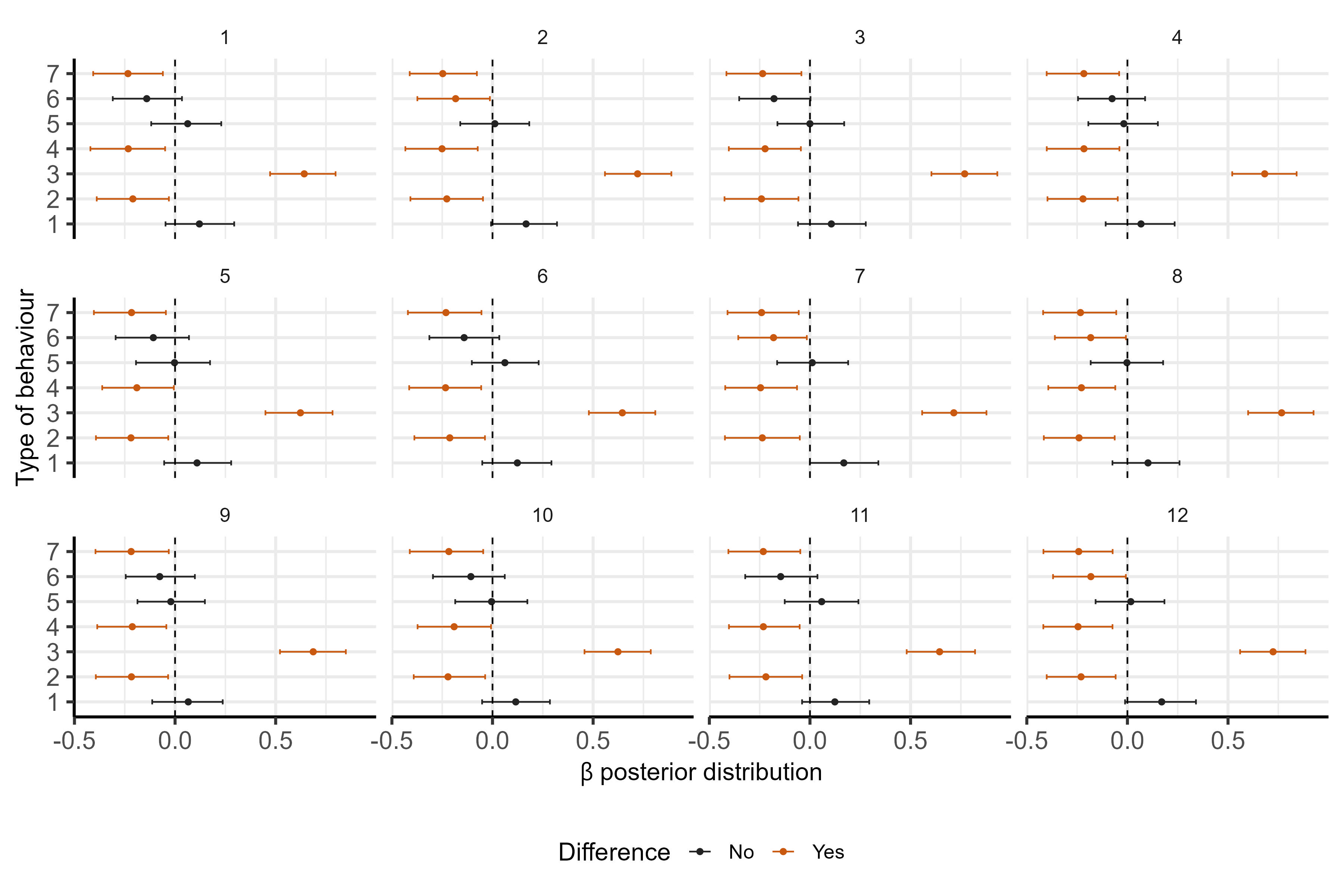}
 \caption{$95\%$ credibility intervals for the estimates of $\beta$, derived from the posterior distribution, indicating the behaviour of male pigs under the environmental enrichment condition with a branched chain. The data were collected over 12 weeks of experimentation, observing the following behaviours: (1) Aggressive, (2) Feeding, (3) Calm, (4) Animal interaction, (5) Environmental interaction, (6) Object interaction, and (7) Locomotion.}
 \label{macho_corrente}
\end{figure}

\begin{figure}[H]
\centering
\includegraphics[width=1.0\textwidth]{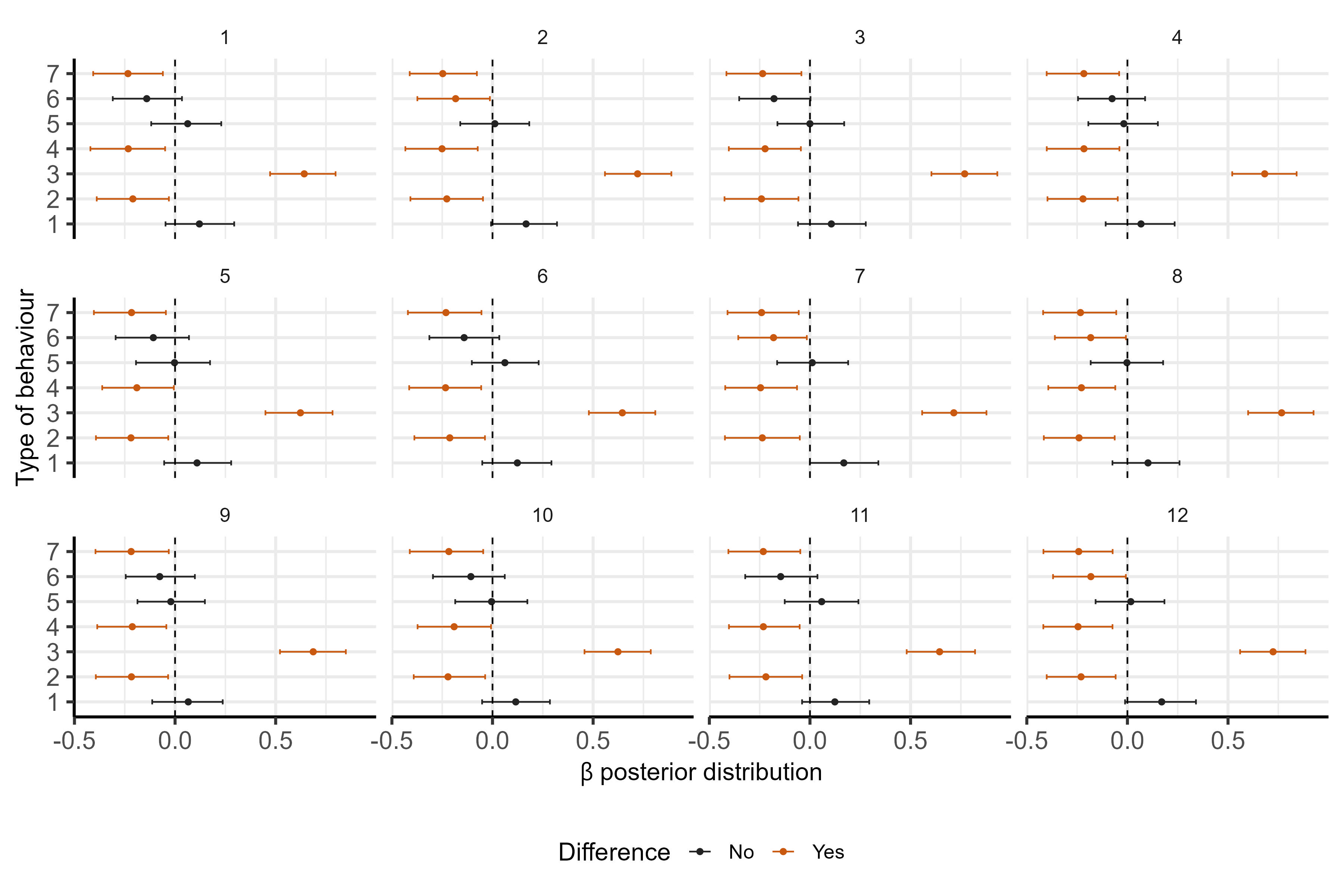}
 \caption{$95\%$ credibility intervals for the estimates of $\beta$, derived from the posterior distribution, indicating the behaviour of male pigs without an environmental enrichment condition. The data were collected over 12 weeks of experimentation, observing the following behaviours: (1) Aggressive, (2) Feeding, (3) Calm, (4) Animal interaction, (5) Environmental interaction, (6) Object interaction, and (7) Locomotion.}
 \label{macho_SE}
\end{figure}

The behaviour of male pigs (Figures \ref{macho_corda}, \ref{macho_corrente}, and \ref{macho_SE}), the $95\% $ credibility intervals indicate that when subjected to environmental enrichment with a sisal rope, the behaviour ``Calm'' was predominant during the first six weeks of the experiment compared to the other behaviours. ``Object Interaction'' was statistically significant between weeks 3 and 8, except for week 6, although the estimates suggest a low preference compared to other behaviours. Meanwhile, `` Environment Interaction'' was significant between weeks 3 and 11, except for week 10. In the final week of the experiment, the behaviour ``Feeding'' was the only one that remained significant.  

For male pigs exposed to environmental enrichment with a branched chain, a greater preference for the ``Calm'' behaviour was also observed throughout the experiment. However, an interesting aspect was these animals' interest in ``Object interaction'' in the final week of the experiment. Nevertheless, compared to enrichment with a sisal rope, ``Object interaction'' was less pronounced when the branched chain was used. Finally, for male pigs that were not subjected to any type of environmental enrichment, the most preferred behaviour was ``Calm''.

The results indicate that environmental enrichment influenced the behaviour of the pigs, with variations between the types of enrichment to which they were subjected. The sisal rope reduced the ``Aggressive'' behaviour, particularly among the females, while also stimulating behaviours such as ``Environment Interaction'' and ``Object Interaction''. In contrast, the branched chain had a lesser effect on the types of behaviours adopted by the animals and was associated with ``Aggressive'' behaviour, especially in the females. The absence of enrichment resulted in the predominance of ``Aggressive'' behaviour among the females and less behavioural diversity in the males, suggesting potential impacts on animal welfare.

\newpage

\section{Conclusions}

\noindent

This study explored the use of Bayesian hierarchical models as an alternative to the analysis of nominal polytomous data in longitudinal studies, addressing the issues of maximum likelihood estimation often associated with overparameterisation. Therefore, this study did not aim to compare with classical statistical methods, especially because in our study, there is no convergence due to the excess of parameters.

The process was illustrated with an application in the field of Animal Science involving seven response categories, and the results indicated that Bayesian hierarchical models provide robust estimates and consistent interpretations in relation to the experiment. Furthermore, when applied in animal science, these models revealed valuable insights into animal behaviour, with practical implications for management and welfare. Additionally, the results confirm that Bayesian hierarchical models offer robust estimates and consistent interpretations, demonstrating their efficacy in the analysis of longitudinal nominal polytomous data while addressing convergence issues.

Finally, it is concluded that the use of Bayesian hierarchical models for nominal polytomous data in future research is recommended, not only in animal science but also in other areas of agricultural and biological sciences. This approach, combined with the development of accessible computational tools, has the potential to significantly enhance the analysis of complex data in longitudinal contexts, enriching empirical research across various scientific domains. For future work, residual analysis and diagnostic techniques should be explored for these classes of models.

\subsection*{Acknowledgments}
This publication has emanated from research conducted with the financial support of the Coordination for the Improvement of Higher Education Personnel (CAPES), process number 88882.378344/2019-01, the National Council for Scientific and Technological Development (CNPq) and Taighde Éireann – Research Ireland under Grant number 18/CRT/6049. The opinions, findings, and conclusions or recommendations expressed in this material are those of the authors and do not necessarily reflect the views of the funding agencies.

\section{Declarations}

\textbf{Competing interests} The authors declare no conflicts of interest.

\textbf{Authors’ contributions} Maria Letícia Salvador and Idemauro Antonio Rodrigues de Lara. {\bf Data curation}:  Mariana Coelly Modesto Santos Tavares and Iran José de Oliveria Silva. {\bf Formal analysis}: Maria Letícia Salvador and Gabriel Rodrigues Palma. {\bf Funding acquisition}: Science Foundation Ireland  and Coordenação de Aperfeiçoamento de Pessoal de Nível Superior. {\bf Investigation}: Mariana Coelly Modesto Santos Tavares and Iran José de Oliveria Silva.  {\bf Methodology}: Maria Letícia Salvador, Gabriel Rodrigues Palma and Idemauro Antonio Rodrigues de Lara. {\bf Project administration}: Iran José de Oliveria and Idemauro Antonio Rodrigues de Lara. {\bf Software}: Gabriel Rodrigues Palma. {\bf Resources}: Mariana Coelly Modesto Santos Tavares and Iran José de Oliveria. {\bf Supervision}: Idemauro Antonio Rodrigues de Lara and Iran José de Oliveria. {\bf Validation}: {\bf Visualization}: {\bf Writing - original draft}: Maria Letícia Salvador, Gabriel Rodrigues Palma, Mariana Coelly Modesto Santos Tavares,  Iran José de Oliveria and Idemauro Antonio Rodrigues de Lara.  {\bf Writing - review and editing}: Iran José de Oliveria and Idemauro Antonio Rodrigues de Lara. 

\bibliographystyle{apalike}
\bibliography{ref}

\end{document}